\documentclass{moriond}

\usepackage{amsmath}
\usepackage{mathtools}
\usepackage[separate-uncertainty=true]{siunitx}
\usepackage{csquotes}

\def\Journal#1#2#3#4{{#1} {\bf #2}, #3 (#4)}

\def\PRL{\em Phys. Rev. Lett.}

\def\NJP{\em New J. Phys.}
\def\PTP{{\em Prog.\ Phys.\ }}

\def\be{\begin{equation}}
\def\ee{\end{equation}}
\def\bea{\begin{eqnarray}}
\def\eea{\end{eqnarray}}
\def\CP{C\!P}

\def\Bs{B^0_s}
\def\Bsb{\kern0.18em\overline{\kern -0.18em B}{}^0_s}
\def\Bd{B^0_d}
\def\Bdb{\kern0.18em\overline{\kern -0.18em B}{}^0_d}
\def\Dsm{D_s^-}
\def\Dsp{D_s^+}
\def\Dsmp{D_s^\mp}
\def\pip{\pi^+}
\def\pim{\pi^-}
\def\dms{\Delta m_s}
\def\dmd{\Delta m_d}

\DeclareSIUnit\invps{\pico\second^{-1}}

\newcommand{\Photo}{}

\begin{document}
\vspace*{4cm}
\title{PRECISION MEASUREMENT OF THE $\Bs-\Bsb$ OSCILLATION FREQUENCY
$\dms$ WITH $\Bs\to\Dsm\pip$ DECAYS AT LHCB}

\author{KEVIN HEINICKE\\on behalf of the LHCb collaboration}

\address{Technische Universität Dortmund, Fakultät Physik\\
Otto-Hahn-Str.~4a, 44227 Dortmund, Germany}

\maketitle\abstracts{
    Decays of $\Bs\to\Dsm\pip$ allow to determine the oscillation frequency
    $\dms$ between the $\Bs$ particle and $\Bsb$ antiparticle states with high
    precision.
    It is a crucial input to constrain the CKM matrix and a manifestation
    of the quantum nature of physics.
    A new measurement of this frequency is presented, using a dataset
    corresponding to \SI{6}{fb^{-1}} of $pp$ collisions, recorded by LHCb at
    a centre-of-mass energy of \SI{13}{\tera\electronvolt}.
    The oscillation frequency is determined to be $\dms =
    \SI[parse-numbers=false]{17.7683\pm 0.0051(\text{stat.})\pm0.0032(\text{syst.})}{\invps}$
    and a combination with previous LHCb measurements yields $\dms^\text{LHCb}
    = \SI{17.7656(57)}{ps^{-1}}$.
}

\section{Introduction}
In nature, neutral mesons such as $\Bs$ mix with their antiparticle, $\Bsb$.
This effect is well-described in the Schrödinger image, and is a beautiful
example of the quantum nature of particles.
Within the Standard Model, the mixing process is governed by the weak
interaction and is only possible due to the non-zero off-diagonal elements of
the Cabibbo-Kobayashi-Maskawa (CKM) quark mixing matrix
\cite{Cabibbo,KobayashiMaskawa}.
A precise measurement of the oscillation frequency $\dms$ adds a
strict constraint on the CKM triangle apex \cite{pdg}, that is even more
powerful when combined with the measurement of the $\Bd-\Bdb$ oscillation
frequency, $\dmd$.
Moreover the value of $\dms$ is a crucial input parameter for measurements of
$\CP$ violation in decays such as $\Bs\to\Dsmp K^\pm$.

The generally high precision of this measurement is possible due to the
large ratio of the oscillation frequency and the lifetime of $\Bs$ mesons.
Given the current world averages of these parameters, $\Bs$ and $\Bsb$
oscillate on average approximately four times per mean lifetime.
In addition to being a test of the Standard Model, a measurement of $\dms$
serves as a benchmark of the detector performances and data analysis methods.

Mixing of $\Bs$ mesons was first observed by the CDF collaboration \cite{cdf}.
The first measurement of $\dms$ carried out by the LHCb collaboration
\cite{lhcb_dms_2011} used a dataset corresponding to an integrated luminosity
of $\mathcal{L}_\text{int} = \SI{1}{fb^{-1}}$.
More recently, the LHCb collaboration measured $\dms$
using data collected during Run~1 and Run~2 with decays of
$\Bs\to\Dsm\pip\pim\pip$ \cite{lhcb_dms_ds3pi}.
The measurement presented in this document is based on the full Run~2 data
sample, corresponding to an integrated luminosity of \SI{6}{fb^{-1}} which
contains a total number of approximately \num{380} thousand signal decays of
$\Bs\to\Dsm\pip$.
This data sample allows for the most precise single measurement of $\dms$ as of
today.
Moreover, all measurements of $\dms$ that have been performed by LHCb are
combined.

The LHCb detector is a single-arm forward spectrometer covering the
pseudorapidity range of $\num{2} < \eta < \num{5}$, designed to study hadrons
containing $b$ or $c$ quarks.
It includes a high-precision tracking and vertex detection system, which
consists of a silicon-strip vertex detector that surrounds the $pp$ interaction
region, a large-area silicon-strip detector located upstream of a dipole
magnet, and three tracking stations placed downstream of the magnet that each
feature an inner silicon-strip detector and an outer straw drift tube detector.
A fundamental requirement for the presented analysis is LHCb’s excellent
decay-time resolution of about \SI{45}{fs}, which allows to measure the fast
$\Bs$ oscillation.
Another crucial ingredient for the analysis is the distinction between
kaons, pions, and protons, which the LHCb detector accomplishes with two
ring-imaging Cherenkov detectors that are part of the particle identification
(PID) system.
LHCb’s trigger system features a hardware and a software stage, which together
reduce the event rate to about \SI{10}{k\hertz}.
More details about the LHCb detector can be found elsewhere
\cite{lhcb_detector}.

\section{Mixing}
The neutral $\Bs$ meson system can be described as a superposition of the two
flavour eigenstates, $|\Bs\rangle$ and $|\Bsb\rangle$.
The time-evolution of the system is obtained by solving a Schrödinger
equation in the heavy and light mass eigenstates $|B_H\rangle$ and
$|B_L\rangle$, with masses $m_{H,L}$ and decay widths $\Gamma_{H,L}$,
respectively \cite{nierste}.
The solution for the flavour eigenstates can then be written in terms of the
decay width $\Gamma_s = (\Gamma_H + \Gamma_L) / 2$, the decay width difference
$\Delta\Gamma_s = \Gamma_L - \Gamma_H$, the mass $m_s = (m_H + m_L) / 2$ and
the mass difference $\dms = m_H - m_L$.
It provides a prediction for the decay-time dependent decay rates
$\Gamma(\Bs(t)\to f)$ of the $\Bs$ mesons into an exclusive final state $f$.

In the special case of flavour-specific decays, such as $\Bs\to\Dsm\pip$,
direct decays from the same initial flavour into a $\CP$ conjugate final state
-- $\Bs\to\Dsp\pim$ in this case -- are heavily suppressed.
Therefore, the electric charge of the final state particles unambiguously
determines the flavour at decay.
Additionally, information about the initial flavour can be obtained through
flavour tagging techniques which exploit the decays of hadronization partners
of the signal $\Bs$ meson.
Knowing both the initial and final flavour of the decays allows to distinguish
all possible decay rates.

As of today, direct and indirect $\CP$ violation can be neglected for decays
of $\Bs\to\Dsm\pip$, such that the decay-time dependent decay rate of the
process reads like
\begin{equation}
\Gamma(\Bs(t)\to\Dsm\pip) = \frac{1}{2}\mathcal{N}e^{-t\Gamma_s}\Bigl[\cos(\dms t) + \cosh\Bigl(\frac{\Delta\Gamma_s t}{2}\Bigr)\Bigr]\,,
\end{equation}
with a normalization factor $\mathcal{N}$.
The $\CP$ conjugate process is described with the same formula, except for a
potentially different normalization factor.
Both of these processes are henceforth referred to as unmixed decays.
The mixed decays, $\Bs\to\Bsb\to\Dsp\pim$ and the charge conjugate process,
only differ by a relative sign in front of the trigonometric term.
Since the direct decay is suppressed in flavour specific decays, they are
labelled as $\Bsb\to\Dsm\pip$.

\section{Measurement}
The measurement is performed on a dataset that corresponds to an integrated
luminosity of \SI{6}{fb^{-1}}.
It has been recorded at a centre-of-mass energy of $\sqrt{s} = \SI{13}{TeV}$
with the LHCb detector.
To improve the signal-to-noise ratio, an event selection is applied, making use
of different kinematic variables of the data sample, as well as the excellent
particle identification information that is obtained from LHCb's RICH
detectors.
In addition, a boosted decision tree (BDT) is used to greatly reduce the number of
random track combinations in the data sample.

\begin{figure}[t]
    \centering
    \includegraphics[width=1.0\linewidth]{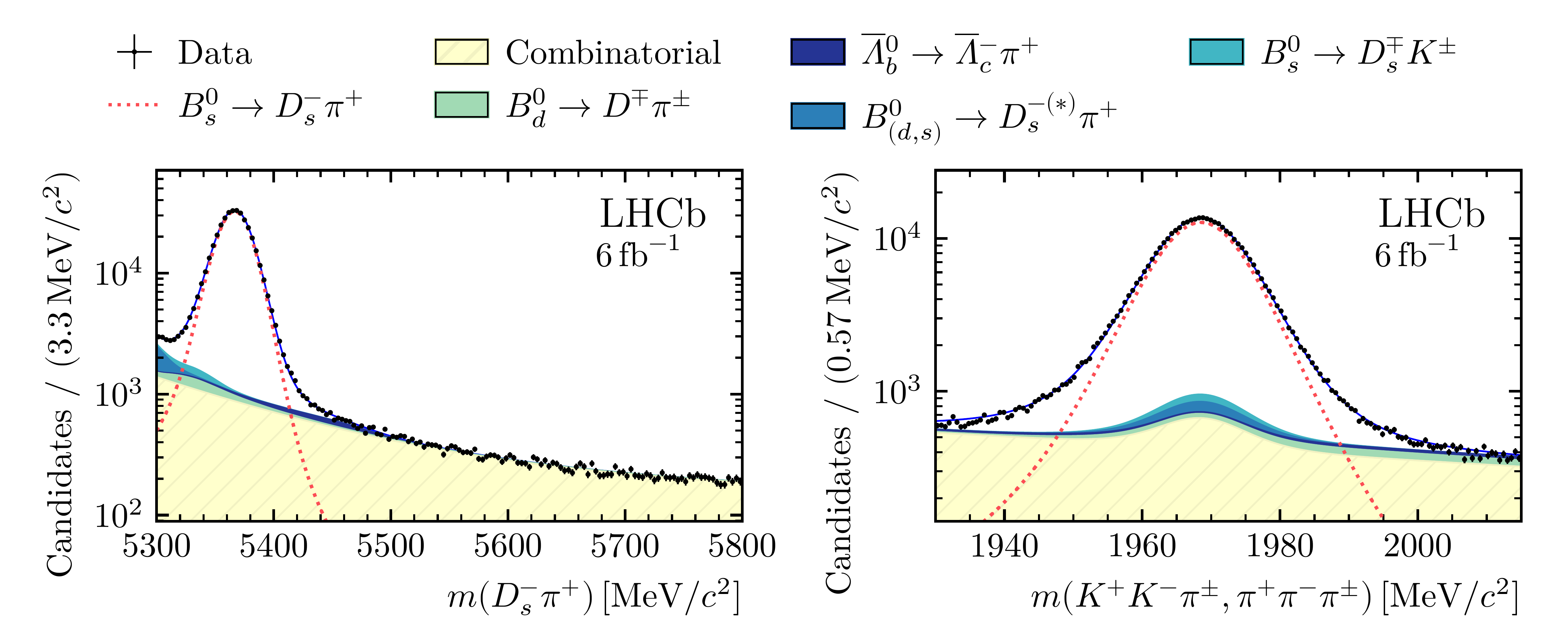}
    \caption{
        Distributions of the invariant $\Bs$ (left) and $\Dsm$ (right) masses,
        with the maximum likelihood fit result superimposed.
        The fit allows to statistically extract the decay-time distribution of
        a pure signal sample.
    }
    \label{fig:mass}
\end{figure}

The good mass resolution that is obtained after these selections is exploited
to statistically extract the signal component of the sample via the sPlot
method \cite{sPlot}.
Both, the invariant $\Bs$ and invariant $\Dsm$ mass distributions are therefore
fitted simultaneously using a maximum-likelihood fit, as shown in
fig.~\ref{fig:mass}.
The event weights are used to extract a decay-time distribution corresponding
to \SI{\sim380}{k} signal decays is obtained.

The initial flavour is measured by a combination of six flavour tagging
algorithms.
Since not all events can be tagged and the algorithms have a certain fraction
of wrongly tagged events, this results in an effective tagging efficiency of
\SI{\sim6}{\percent}.
The decay-time uncertainty is described with a Gaussian resolution function and
calibrated using $\Dsm\pip$ pairs originating from the $pp$ interaction region.
Additionally, detector effects and the BDT selection introduce a decay-time
dependent efficiency, which is modelled with a set of cubic spline functions.

\begin{figure}[b]
    \centering
    \includegraphics[width=0.5\linewidth]{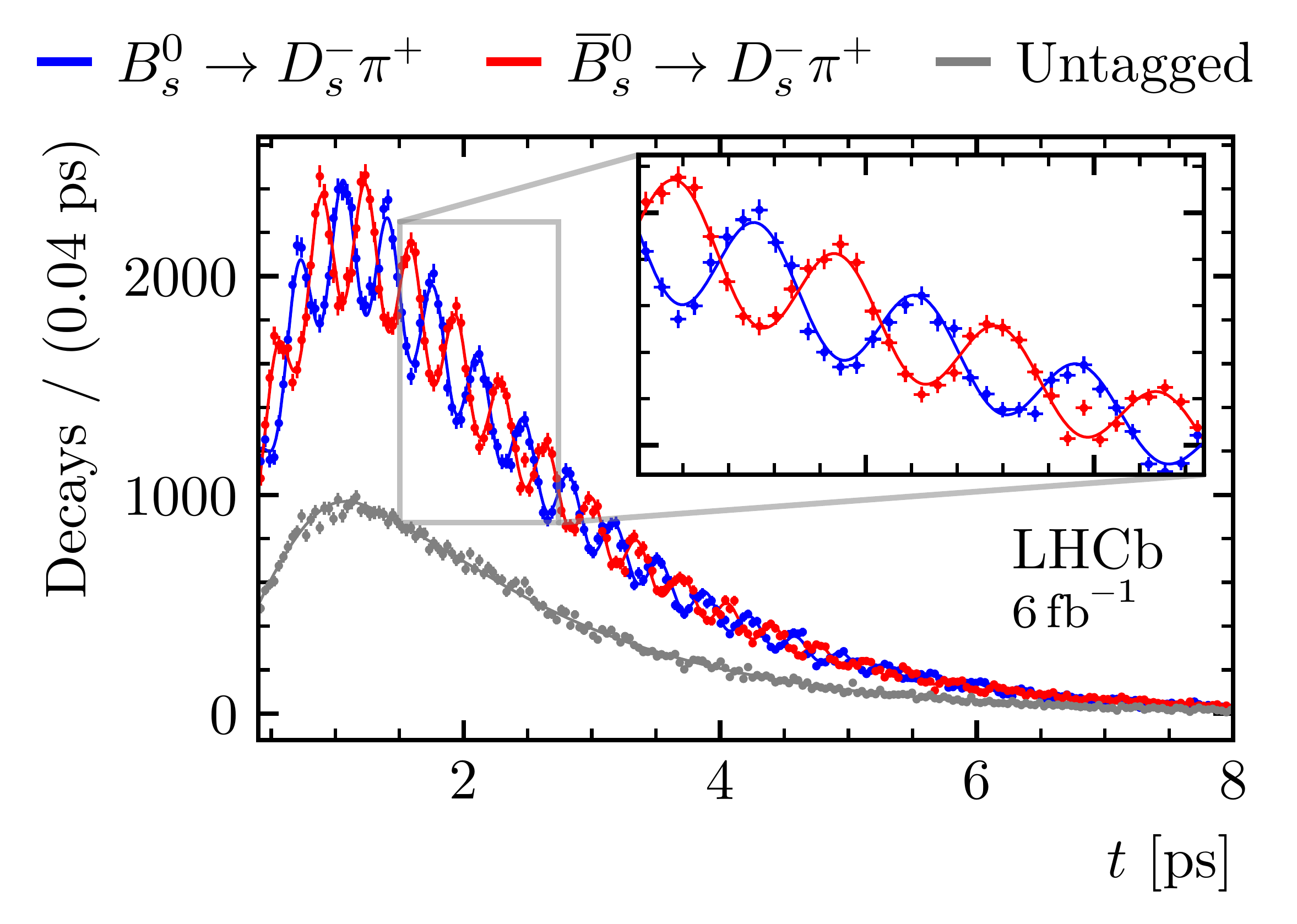}
    \caption{
        Decay-time distribution of unmixed (blue), mixed (red) and untagged
        (grey) signal decay.
        The data points are nicely described with a cosine oscillation across
        the full range.
    }
    \label{fig:osc}
\end{figure}

\begin{figure}[h]
    \centering
    \includegraphics[width=1.0\linewidth]{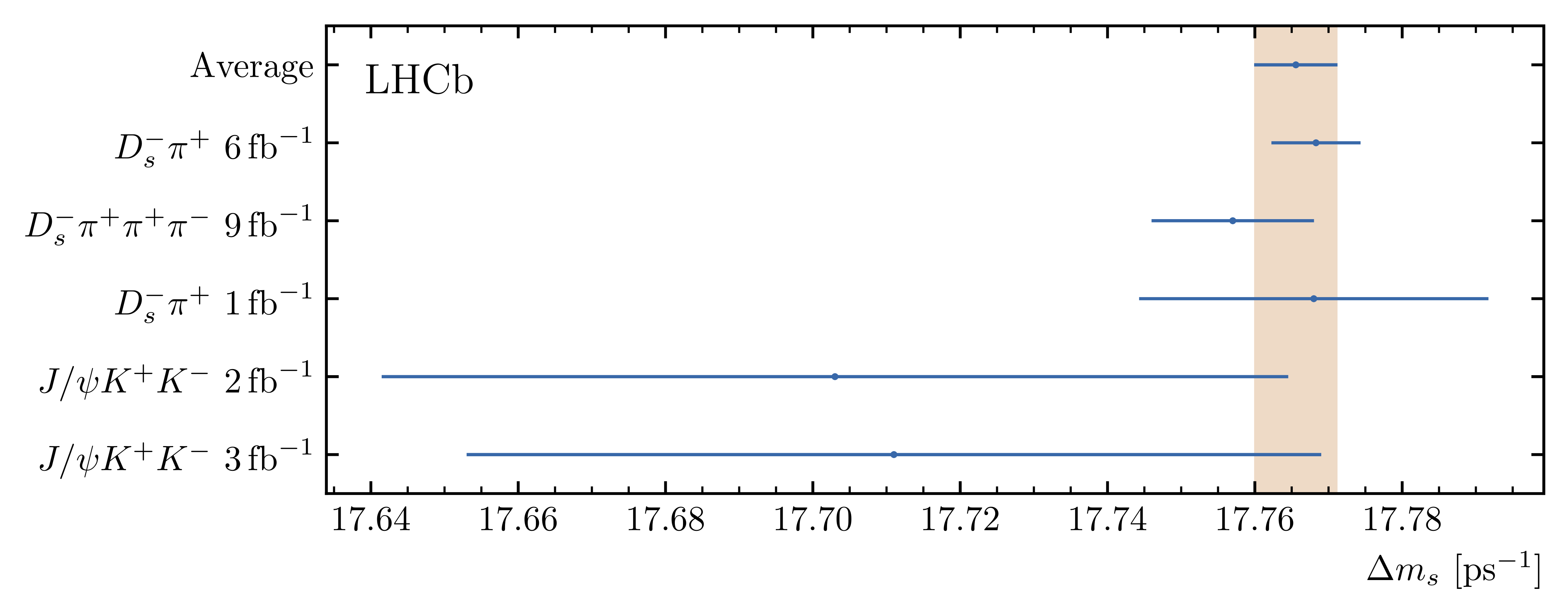}
    \caption{
        Combination of several LHCb measurements of $\dms$, and the individual
        results.
        The average is clearly dominated by the most precise measurement
        presented here.
    }
    \label{fig:comb}
\end{figure}

The resulting decay-time distribution, split into unmixed decays
$\Bs\to\Dsm\pip$, mixed decays $\Bsb\to\Dsm\pip$, and untagged decays is shown
in fig.~\ref{fig:osc}.
The measurement yields
\begin{equation}
    \dms = \SI[parse-numbers=false]{(17.7683\pm0.0051\pm0.0032)}{ps^{-1}}\,,
\end{equation}
where the first uncertainty is statistical and the second is systematic.
Due to the high precision of this measurement, the systematic uncertainties
need to be controlled to unprecedented levels.
The dominant systematic uncertainty originates from the imperfect knowledge of
the detector alignment.
Additionally, even well-established data analysis tools such as the sPlot
method significantly contributes to the uncertainty budget.
Therefore, the measurement of $\dms$ provides an excellent benchmark for both
detector performance and data analysis tools.
\clearpage
Ultimately, the presented measurement of $\dms$ is combined with other
LHCb measurements of the same parameter
\cite{lhcb_dms_2011,lhcb_dms_ds3pi,lhcb_jpsikk1516,lhcb_jpsikkrun1}, as shown
in fig.~\ref{fig:comb}, taking correlated systematic uncertainties into
account.
It results in the world's most precise determination of the $\Bs$--$\Bsb$
oscillation frequency as of today, $\dms^\text{LHCb} =
\SI{17.7656(57)}{ps^{-1}}$.

\section{Conclusion}
The results presented here include the most precise single measurement of the
oscillation frequency $\dms =
\SI[parse-numbers=false]{(17.7683\pm0.0051\pm0.0032)}{ps^{-1}}$, as well as the
most precise single experiment determination of the parameter,
$\dms^\text{LHCb} = \SI{17.7656(57)}{ps^{-1}}$.
It is a crucial ingredient to CKM triangle measurements and a good benchmark
for future detector and analysis developments.
More than that, it is a beautiful example of quantum effects in particle
physics.

\section*{Acknowledgements}
The author has been supported by the German Federal Ministry of Education and
Research (BMBF).

\end{document}